# Analyzing Common Social and Physical Features of Flash-Flood Vulnerability Hotspots in Urban Areas

Allison Clarke, allison.clarke@tamu.edu, Natalie Coleman, ncoleman@tamu.edu, Dr. Ali Mostafavi, amostafavi@civil.tamu.edu

**Abstract**

Flash flooding events, with their intense and sudden nature, present unique challenges for disaster researchers and emergency planners. To quantify the extent to which hotspots of flash flooding share similar social and physical features, the research uses community-scale crowdsourced data and k-means clustering. Crowdsourced data offers the potential to allocate limited resources, to improve spatial understanding, and to minimize the future effects of natural hazards. The research evaluates the impacts of Tropical Storm Imelda on Houston Metropolitan and Hurricane Ida on New York City. It develops a combined flash flood impact index based on FEMA claims, 311 calls, and Waze traffic reports which is able to capture a combination of crowdsourced data for the societal impact of flash flooding. In addition, k-means clustering offers an essential tool for evaluating attributes associated with flash flooding events by grouping data into k-number clusters of features. Thus, k-means clustering evaluates the significance of a community's socio-demographic, social capital, and physical features to the combined flood impact index. To ensure accessibility and replicability to different types of communities, our research uses publicly available datasets to understand how socio-demographic data, social capital, and physical connectivity and development affect flash flood resilience. The findings reveal the intricate relationships associated with flash flooding impact as a combination of socio-demographic and physical features. For instance, the cluster with the highest scores of the flash flood impact index from Tropical Storm Imelda had the highest percentage of minority population and lower income while the cluster with the second highest score flash flood impact index had the highest percentage of impervious surface and number of POIs.

## Introduction

Flash flooding events cause detrimental effects to human health, infrastructure systems, and the overall well-being of communities in the United States (US). In fact, flash floods are the largest cause of weather-related property damage with an estimated cost of $49 billion dollars between 2017 and 2021 (Davis, 2022). For the past decade in the US, the severity and intensity of flash flooding have been increasing, and this trend is expected to continue (Alipour et al., 2020). Particularly at risk are urban areas, where high levels of impervious surfaces and frequently overburdened stormwater systems heighten vulnerability. Flash flooding impacts are further compounded by dense construction and increased population.

The rapid onset of flash floods typically occurs within six hours of heavy rainfall(National Oceanic and Atmospheric Administration (NOAA)). This further constrains the time available for effective emergency response and decision-making, which are crucial for safeguarding people and infrastructure. Conventional flood risk indicators, such as the 100-year flood map, often inadequately represent the dynamic and immediate nature of flash flooding risks (Mobley and Blessing, 2022). Moreover, the current placements of flood gauges may not adequately capture the full extent of impacts in critical areas (Farahmand et al., 2022). This evolving and urgent context establishes the importance of understanding the characteristics which define flash flood hotspots to better prepare and respond to flash flood risks.

While recognizing the need for improved understanding and response to flash flood risks, particularly in urban areas, it is important to evaluate the current modeling approaches to these natural hazards. The majority of flash flooding models predominantly focus on hydrological features such as rainfall intensity and soil moisture conditions (Henao Salgado and Zambrano Nájera, 2022; Zhang et al., 2022; Mohtar et al., 2020). However, research has demonstrated that a plethora of variables can influence the likelihood and consequential impacts of flash flooding (Abdel-Mooty et al., 2021; Fernandez et al., 2016; Liu et al., 2024). To accurately quantify the impact of flash flood events, it is crucial to consider both the physical and social impacts on communities (Špitalar et al., 2014; Karagiorgos et al., 2016).

Flash flooding is influenced by physical factors such as land development and rapid urbanization as well as the proximity and accessibility of resources within a region. Rapid urbanization is linked to an increase in both the frequency and severity of flash flooding (Blum et al., 2020; Feng et al., 2021). This escalation is largely attributed to an increase in impervious surfaces and the channelization of streams, which alter the natural hydrological system and escalate flash flood risk (Cutter Susan et al., 2018). Additionally, urbanization often induces the urban heat island (UHI) effect, which can alter precipitation patterns (Lin et al., 2011). Moreover, areas with high road densities are often marked by significant modifications to the natural landscape, which can disrupt water storage by floodplains and affect the movement of natural sediments (Rahman et al., 2021). Physical resilience is also strongly influenced by accessibility and proximity of resources within a community (Patrascu et al., 2023; Li et al., 2023; Yuan et al., 2023) .

In addition to physical attributes which affect flood risk, the impact of flash floods is not uniform across different socio-demographic groups, with low-income households, racial-ethnic minorities, and the elderly often disproportionately affected (Faber, 2015; Rufat et al., 2015; Tate et al., 2021). These social inequities shape the vulnerability of various groups to flash flood risk, affecting their ability to recover and respond to flood hazards (Sanders et al., 2023). The relationship between socioeconomic status and high hazard exposure offers a potential source of environmental inequality, which poses a critical challenge within flood risk assessment and mitigation practices (Chakraborty et al., 2020). The risk and impact of flash floods on digitally invisible populations - groups underrepresented or absent in digital data and online platforms - may be overlooked or misunderstood through emerging data collection methods (Longo et al., 2017). These populations are even more likely to include lower income, elderly, or rural populations due to inaccessibility to mobile phones, unfamiliarity with digital platforms, or lack of financial resources, respectively. Additionally, social capital networks have also been associated with mitigating the effects of flash flooding (Azad and Pritchard, 2023; Działek et al., 2014). Disparities in social capital may limit the ability of different groups to effectively prepare for and respond to flash flood risk, exacerbating inequities of flood resilience. Therefore, it is important to consider these influential physical and social drivers to better characterize the hazard-impact on communities and bridge the gap between engineering and social science solutions(Robertson, 2023).

Regarding concerns of rapid urbanization, community connectedness, and digitally invisible populations, existing studies have shown how community-scale data can be used to evaluate and identify flood risk to help people avoid high-impact areas and to direct resource allocation (Ramadhan; Supriyadi et al., 2018). Crowdsourced data can provide a quick-time accurate depiction of hazards (Farahmand et al., 2022; Yuan and Liu, 2018). In this context, crowdsourced data, including FEMA flood claims, 311 calls, and Waze traffic reports, offer a valuable analysis tool. FEMA flood claims can be used to estimate the level of damage of flooding events(Mobley et al., 2021). Additionally, the reporting of 311 calls (Rainey et al., 2021; Faber, 2015) and Waze traffic data (Lowrie et al., 2022; Praharaj et al., 2021) have also been proven to be reliable in other assessment studies. Therefore, this study utilizes crowdsourced data to capture the societal impacts of flash flooding. Such data can reveal relationships between multiple variables and examine the intricate relationships to flood impact. By combining the FEMA flood claims, 311 calls, and Waze reports, the research aims to mitigate the data biases related to systemic exclusions, unfamiliarity, and access to internet services, respectively. The research proposes the creation of a combined flood impact index using the three data sets to measure the flooding impact of Tropical Storm Imelda in Harris County, TX and Hurricane Ida in New York City.

  The objective of the study is to integrate the physical, socio-demographic, and social capital dimensions of communities to better understand the shared social and physical characteristics of flash flood hotspots. Using a cross-analysis of flood impact, the study aims to highlight patterns in social and physical characteristics that may reveal inequities in flash flood exposure and to identify urban development features which exacerbate flood risk. K-means clustering analysis develops distinct groups of comparison for interpreting and characterizing the associated factors to the flood impact index (Lu and Zhou; Xu et al., 2018; Ramadhan; Supriyadi et al., 2018). The guiding research question is "to what extent, do social and physical features explain flood inequities?" First, identifying similar social features in high-impact clusters may reveal underlying inequalities in flash flood exposure. Second, shared physical characteristics in these areas could indicate common urban development features that contribute to heightened flood risk. It will also draw attention to underrepresented and digitally invisible populations through better data representation in crowdsourced data and integrated multi-variable analysis in k-means clustering. The research will ensure use of publicly available datasets, such as US census data, road density, impervious surface, POI development, and social capital measurements to ensure the replicability by involved stakeholders. The findings may be used to contribute to the development of more equitable and effective flood risk mitigation strategies. If social and physical inequalities are uncovered, data may be used to drive urban planning decisions and to direct attention to vulnerable populations.

**Literature Review**
  The literature review begins by discussing the importance of social capital in building resilience against flash floods. Next, it investigates the relationships between physical and socio-demographic characteristics and flash flood vulnerability. Finally, it emphasizes the value of integrating different sources of crowdsourced data to assess flash flood risk.
  **Understanding the Effects of Urbanization on Flash Flood Risk:** The increasing urbanization and development of communities has exacerbated the effects of flooding (Sjöstrand, 2022). Impervious surfaces can lead to higher surface runoff (Sohn et al., 2020) prior to flash flooding events as well as higher river discharge rates (Feng et al., 2021). Based on US stream gauge data, a percentage increase of impervious cover is associated with a 3-5% increase of flood magnitude (Blum et al., 2020). Additionally, land use changes result in "changing physical properties such as albedo, surface roughness, evapotranspiration and energy flux," which may create urban heat islands (UHI) (Lin et al., 2011). According to Rachma and Lin (2023), urban heat islands (UHIs) exhibit a positive correlation over time with rainstorms, attributable to increased atmospheric instability and enhanced thunderstorm development. Based on their analysis of the Taipei basin in Taiwan, the number of thunderstorms in Taipei has risen by 25% across the past two decades and has become more frequent in metropolitan areas, indicating an increased risk of flash floods in urban areas.

Another important component of flood resilience is the capability of the built environment to "resist and rapidly recover from disruptive events"(McAllister and McAllister, 2013).Patrascu et al. (2023) supports this sentiment by asserting that facility distribution and access are vital during and after disasters. Consequently, communities with a high concentration of points of interest (POI) may indicate increased ability to respond to and recover from hazards. POIs include places like hospitals, fire stations, schools, grocery stores, utility services, transportation hubs, and cultural centers. On the other hand, high levels of POIs in close proximity may signal increased development and urbanization, which brings the previously mentioned effects of flash flooding. For instance, through a flash flood simulation, Feng et al. (2021) found that a 100% impervious surface area doubles the affected areas and maximum depths of flooding compared to a 0% impervious surface condition. Therefore, land use distribution, particularly in terms of urban development and impervious surfaces, greatly affects the extent of flash flooding.

Shifting focus from the broader effects of urbanization, Rahman et al. (2021) found that road density has a significant correlation (0.496 correlation coefficient) with the occurrence of floods. This suggests that the distribution of roads is directly related to flooding. Similarly, Jones et al. (2000) assert that road networks are "hydrologically connected to the stream networks" based on evidence that roadways may increase flood peaks, modify flood routing patterns, and cause more frequent debris flow disturbances. As an example, Napieralski and Giroux (2019) observed that approximately 72 percentage of stream channels in the Rouge River watershed, a highly urbanized watershed in southeast Michigan, were classified as "proximally affected by the road network." This classification reflects changes in watershed size, shape, and localized runoff patterns, as road networks often form localized topographic highs and alter watershed boundaries, leading to increased alignment between stream and road networks. These findings signal that the construction and layout of road networks can have varied and significant hydrological impacts on different watersheds, which can influence the frequency and intensity of floods.

**Associating Socio-demographic Characteristics to Flash Flood Vulnerability:** Certain socio-demographic populations are often overburdened by the effects of urbanization on flash flooding (Chakraborty et al., 2020). These groups often reside in areas that are overexposed to flooding. For example, racial minorities and mobile home residents were overrepresented in a hotspot analysis of flood exposure (Tate et al., 2021). Additionally, lower levels of socioeconomic status are correlated with increased flood-related damages and fatalities (Tellman et al., 2020). Further, a study of Hurricane Katrina in New Orleans showed that "race and socioeconomic status were strongly related to the duration of displacement," noting substantially slower rates of returns for populations with lower education levels and for black individuals (Fussell et al., 2010). Chacowry et al. (2018) also identified lower income levels, lower literacy rates, and higher household sizes as factors associated with slower rates of recovery. Socioeconomic inequalities are compounded by a lack of access to economic, infrastructure, and human resources for vulnerable populations, such as the disabled, elderly, and uneducated, to respond effectively to flood risk (Masozera et al., 2007; Coleman et al., 2020). However, the spatial relationship between flood risk and social groups can be complex. For example, an analysis of the Miami Metropolitan Statistical Area by Chakraborty et al. (2014) found that non-Hispanic black and Hispanic residents were significantly overrepresented in inland flood zones and underrepresented in coastal flood zones, signaling an unequal spatial distribution of flood risk.

Moreover, underrepresented and digitally invisible populations may be overlooked in disaster planning efforts due to a lack of internet connectivity or technology. According to Lee (2024), there are about 13% of people that are 'not online', and these populations disproportionately include people of color, older Americans, people with disabilities, foreign-born populations, and rural residents." While the growth of big data offers significant potential for crisis response and disaster resilience(Overton et al., 2022; Tang et al., 2017), inaccurate or partial data may divert resources from vulnerable areas underrepresented on internet-dependent platforms, like social media (Kraft and Usbeck, 2022). Higher populated areas, like dense urban cities, are also more representative on social media platforms and other crowdsourced platforms and may divert resources from less populated areas(Fan et al., 2020; Esparza et al., 2023). Generalizing human response and behavior based on social networking is inherently unrepresentative and may lead to misleading insights (Mergel et al., 2016), underscoring the need for diversified data sets that are not reliant on broadband access. To address this data gap, the research utilizes 311 data as a method for characterizing flash flood vulnerability (Negri et al.). This quick-time data offers a way to better represent groups who may be overlooked through internet-dependent platforms.

**Measuring Social Capital to Build Flash Flood Resilience**: Previous research has demonstrated that social capital, including bonding, bridging, and linking, can mitigate the negative impacts of flooding (Rustinsyah et al., 2021; Tammar et al., 2020; Aldrich and Meyer, 2015). Social capital principles of self-efficacy and community embeddedness is critical for improving pre-disaster awareness and preparedness, respectively(Hudson et al., 2020) . Individuals are more likely to adopt resilient strategies when influenced by social expectations and deeper community engagement (Lo and Chan, 2017). Communities with strong social capital and mutual trust manage local flood risks more effectively and can develop governance strategies that are more adaptive and responsive to their unique needs (Aldrich, 2010). These networks enable quicker and more effective communication and coordination of resources, essential for supporting vulnerable populations during flooding events. Examples of post-recovery activities include the distribution of emergency relief supplies, infrastructure repair, allocation of financial resources, and collaboration with NGOs. According to Działek et al. (2014), local risk management requires a better understanding of local network capacities and their possible use in local flood risk governance strategies.

**Raising the Representativeness of Crowdsourced Data:**  In many cases, National Flood Insurance Program (NFIP) claims from FEMA are used to assess flood risk and quantify flood damage (Mobley et al., 2021). However, FEMA claims have various limitations, such as whether affected individuals have the time, ability, and specific eligibility to turn in an application for aid as well as whether these insurance applications are ultimately accepted (Emrich et al., 2022; Emrich et al., 2020). Although commonly used to measure flooding impact, FEMA claims alone may be unrepresentative of the true scale and extent of flood impact. Thus, other crowdsourced data provides a supplement to FEMA claims which can be used to quantify the impacts of flooding in urban areas with high-resolution observational data.

One supplemental data source is using 311 call data to show the spatial-temporal distribution of service outages, hazard damage, and recovery process of extreme weather events (Kontokosta and Malik, 2018; Lee et al., 2022; Mobley et al., 2019). For example, Rainey et al. (2021) showed the benefit of using 311 calls to understand the flooding impact during Hurricane Harvey. The study showed that 56% of calls were located outside of FEMA 100-year and 500-year floodplain boundaries, and it supported building damages that were reported by official forensic analysis. Following Superstorm Sandy in New York, 311 calls demonstrated the variation in service distress (Faber, 2015). Certain people may still feel uncomfortable and even afraid to have a 311 call because of governmental distrust (National Academies of Sciences and Medicine, 2019). Lower-income and minority neighborhoods were also found to be less likely to report street condition or "nuisance" issues but did prioritize more serious problems (Kontokosta and Hong, 2021).

Furthermore, Waze reports has the potential to report rapid assessment and mitigation of flash flooding events. Lowrie et al. (2022) also evaluated the usefulness of Waze reports for crowdsourced data. The study found that with timely reporting and relevant information, Waze reports could potentially monitor flood conditions. In a Virginia case study, Praharaj et al. (2021) concluded that 72% of Waze reports were trustworthy. Waze reports were also fairly correlated with well-known flood observations such as documented national flood maps, high watermarks, and low water crossings (Safaei-Moghadam et al., 2023). It had the highest correlation with the likelihood of pluvial flash flooding. We wanted to note that Waze data is accessible to municipalities, states, and countries partnered through the Waze Connected Citizens Program with many open-source platforms and tutorials to process the data (Liu et al., 2023). However, Waze data may have its own limitations in that it is limited to people who can afford and are willing to use Waze app on their smartphones (Lowrie et al., 2022; Praharaj et al., 2021). Given this information, our research uses a combination of FEMA flood claims, 311 calls and Waze reports, to measure the widest scope of flooding impact and account for each data source's own limitations.

**Methodology**

**Case Studies:** On September 18th, 2019, tropical storm Imelda produced heavy rainfall over southeastern Texas. It particularly hit the already flood-stricken city of Houston, where many residents were still recovering from the Tax Day Floods of 2016 and the Hurricane Harvey floods of 2017 (Taylor and Mervosh, 2019). Approximately one thousand vehicles were flooded due to high water on I-10 and throughout the Houston area, and at least 1,700 homes were flooded around Houston (Latto and Berg, 2020)[. The storm was the fifth wettest rainfall event in the United States, dropping more than 41 inches of rain in the most impacted parts of the community (Fernandez, 2019).

On September 1st, Hurricane Ida stalled over New York. The storm traveled to the northeastern parts in the United States, and heavy rainfall was predicted for New York City. Many city and state officials knew about the potential intensity of the storm but could not prepare enough to protect New York City (McKinley et al., 2021a). The hurricane brought historic hourly rainfall, at times more than 3 inches per hour (McKinley et al., 2021b). The aging infrastructure, subways, and basement dwellings were particularly vulnerable to the intense tropical storm. The official death toll was 16 people with most deaths from the Queens area (Marcius and Norman, 2021). The city had previously experienced intense rainfall from Hurricane Henri in 2021 and Hurricane Sandy in 2012. These two flash flooding events demonstrate the quickness and devastation left behind.

**Period of Analysis:** The periods of analysis defined for Tropical Storm Imelda and Hurricane Ida are September 17-20, 2019, and September 1-4, 2021, respectively. These dates are directly before and after the flash flood hit the area of analysis to encapsulate the impact experienced across the geographic region and to represent the delayed impacts of the flash flood. The area of analysis for Tropical Storm Imelda is the Houston Metropolitan Area, and the area of analysis for Hurricane Ida includes the five boroughs of New York City (Bronx, Kings, New York, Queens, and Richmond Counties). All data is analyzed at a census tract level, and census tracts with a population of fewer than 100 people are removed from the data set. These regions primarily represent business centers, shopping districts, parks, and recreational facilities. Their low population count falsely represents impact when data is normalized. One census tract was removed from the Houston data, and 56 census tracts were removed from New York City. Additionally, three census tracts were removed from the New York analysis area due to their high concentration of points-of-interest (POI). These three census tracts are in the 5th Avenue shopping center, Times Square, and Broadway Theater District. For all cases, Houston and New York City data are processed independently.

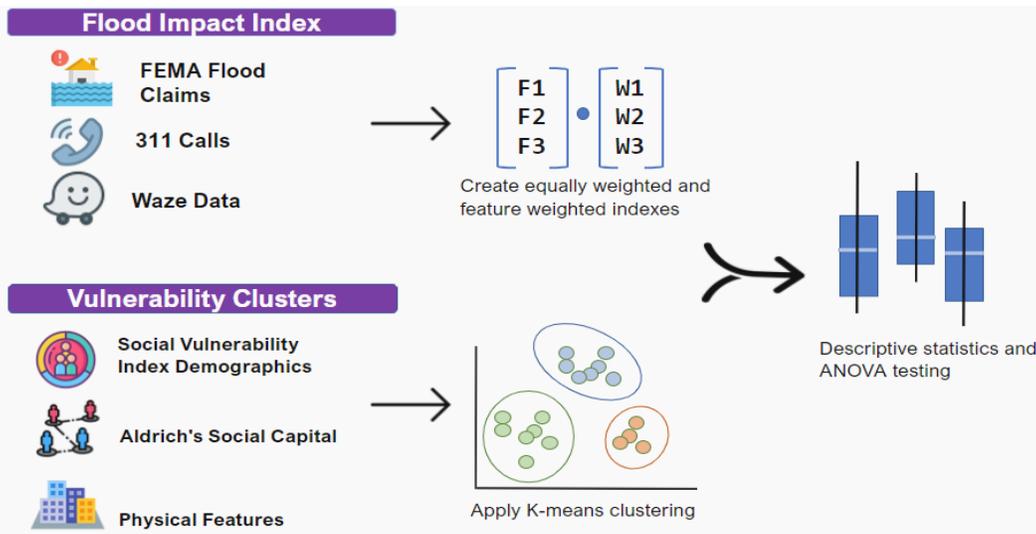

**Fig 1:** Analysis Framework for Creating Flood Impact Index and Vulnerability Clusters

The following sections detail the data sources and methods found in the methodology framework (Fig 1). First, the flood impact index is developed for each census tract to quantify flood impact. Then, vulnerability clusters are established to characterize the social and physical features of each census tract. Lastly, one-way Analysis of Variance (ANOVA) testing is used to assess the statistical significance of social and physical features in identifying areas with high flash flood impact.

**Impact Index Development:** The study's first step is to determine the vulnerability of census tracts according to combined flood impact indices consisting of FEMA claims, 311 calls, and Waze Reports. The impact index is developed by processing and normalizing each data source and then combining the values into a cumulative value between 0 and 1, representing the overall impact. First, all data is normalized using min-max normalization. Then, a weighted sum of values is used to calculate each impact index. One index weights all three impact indicators equally at 33%. The other three indices allow one variable to dominate the overall index, with the dominating variable weighted at 60% and the other two weighted at 20% each. This means if a census tract reported a high number of FEMA flood claims but a low number of 311 calls or Waze-report, the values of each crowdsource data would be applied appropriately based on the weight.

The number of flood-related FEMA claims for each census tract was determined and normalized based on the housing units estimate(Centers for Disease Control and Prevention, 2018). The City of Houston Open Data (Open Data Portal in Houston, 2019) and the City of New York Open Data (2021) (New York City, 2021) are used to obtain 311 calls, which is filtered based on complaint type. Relevant complaint types are selected from increases in call volumes during or after the flash flood event. For Houston, 'Flooding,' 'Drainage,' 'Storm Debris Collection,' and 'Crisis Cleanup' complaints identified flood impact. Differently, 'Storm' and 'Sewer' complaints represent flood impacts in New York City. Next, the number of 311 calls within each census tract was normalized by population. Finally, Waze report data for the period of analysis was obtained directly from Waze and filtered based on the report type (Waze, 2019, 2021). For both data sets, flood-related hazard reports are used to identify flood impact. Given the transportation nature of the Waze traffic data, the number of reports within each census tract normalized based on annual average daily traffic (AADT) data. For Houston, the AADT by census tract is determined using a weighted average of the three closest recorded TxDOT AADT values based on distance from the center of the census tract. For New York City, the AADT sum for each census tract processed through the NYSDOT shapefile data on ArcGIS.

**Socio-Demographic, Social Capital, and Physical Features:** The second step is to characterize each census tract based on social features, social capital, and physical factors. First, data is processed and collected from various publicly available sources. The six sociodemographic features used within this analysis are percentage of minority, per capita income, population density, percentage of persons aged 65 and older, percentage of the civilian noninstitutionalized population with a disability, and percentage of housing in structures with ten or more units (Centers for Disease Control and Prevention, 2018). Three types of social capital are used to define each census tract: bonding, bridging, and linking (Aldrich and Meyer, 2015). Bonding social capital represents the ties between individuals with a relatively high degree of network closure. Bridging social capital represents the ties between individuals which cross social divides or go between social groups. Linking represents networks of trusting relationships between people who interact across explicit, formal, or institutionalized power or authority gradients in society. Points-of-interest (POI) (SafeGraph, 2020), percentage of impervious land cover (Multi-Resolution Land Characteristics and United States Geological Survey (USGS), 2021), and road density(New York City, 2021; Texas Department of Transportation) are used to classify each area's physical features. The number of POIs in each census tract is counted and normalized by population. The percentage of impervious cover data and road density data are processed on ArcGIS Pro. The percentage of impervious cover is calculated using an area-weighted mean, and road density is calculated by summing the total length of roadways within a census tract and dividing this value by the area of the census tract.

**K-means Vulnerability Clusters:** After characterizing the social and physical features of each census tract, k-means clustering is used to determine common features among flash flood hotspots. The k-means clustering algorithm is a machine learning technique that groups data into k-number clusters to identify patterns and similarities within the data. It maximizes between-cluster variance and minimizes within-cluster variance. K-means clustering is an iterative, centroid-based algorithm where each data point is assigned to the closest k-centroid to minimize the Euclidean distance between the data point and its corresponding cluster centroid. Euclidean distance is preferred over other similarity measures because it evaluates the weighted proximity of different objects in a three-dimensional space (Abdel-Mooty et al., 2021).

$$d_e(x, y) = \sqrt{\sum_{i=1}^{n} (q_i - p_i)^2} \qquad (1)$$

, where $q_i$ and $p_i$ represent points in an n-dimensional space to determine the Euclidean distance, $d_e$

One difficulty of using the k-means algorithm is that the optimal number of clusters, k, must be determined in advance. For this analysis, the number of clusters is determined using the elbow method (Nainggolan et al., 2019) , which determines the value of k after which the distortion and inertia start decreasing linearly. Distortion is defined as the average of the squared Euclidean distances from the cluster centers of the respective cluster while inertia is defined as the sum of the squared Euclidean distances of samples to their closest cluster center.

The one-way Analysis of Variance (ANOVA) test is implemented to understand the significance of the difference between variables in a cluster. This test compares the means of each independent cluster to determine if there is statistical evidence that the data values are significantly different. The p-value shows the level of marginal significance when testing the null hypothesis that two or more groups have the same population mean. The f-value is a ratio of explained variance to unexplained variance and is used to determine variance between two populations.

Leveraging crowdsourced data with clustering offers the potential to direct emergency response resources in quick-time and to identify which areas are most highly impacted by a storm event (Liu et al., 2023; Lin et al., 2020). For example, K-means clustering has previously been used to connect the impact of earthquakes, volcanic eruptions, and floods to weather characteristics, such as rainfall and temperature, (Ramadhan), and to categorize areas based on proneness to disasters (Supriyadi et al., 2018). When evaluating flood impact, the features that define an area and their relationship to flood impact are critical information for emergency managers to understand. As the frequency of floods continues to increase, k-means clustering offers an essential tool for evaluating floods' natural and social attributes and improving flood risk management and disaster reduction.

## Results

**Impact Index:** We first created impact indexes from the FEMA flood claims, 311 calls, and Waze reports. To account for the differential influence of each crowdsourced data, four impact indexes with adjusted weights were created. Fig 2. shows the box-plot distribution of the impact values. Many census tracts reported a median value of little to no impact while other census tracts were significantly impacted by the flash flooding events. All data points, especially outliers, were considered comprehensively to understand the impact of flash flooding events. The results indicated that crowdsourcing data may be more influential depending on the location. For example, the Waze-weighted impact index showed the highest scores for Houston while 311-weighted impact index showed the highest scores for New York City. In evaluating the clustering analysis, the overall final scores remained consistent across the impact indexes. This means that if a cluster had the highest impact score for equal-weight impact index, it also had the highest scores for the FEMA Flood Claims-Weight, 311 Calls-Weight, and Waze-Weight. However, the difference in weights is still important for capturing the individual impact of each individual census tract as some tracts may be more inclined to report damage from a FEMA flood claim when compared to 311 call or Waze Report, vice-versa. Fig 3-4 shows the spatial distribution of the impact index values for each census tract. For Fig 3, the most impacted areas are in the North-West of Houston metropolitan, and for Fig. 4, the most impacted areas were in south New York, mostly around Staten Island.

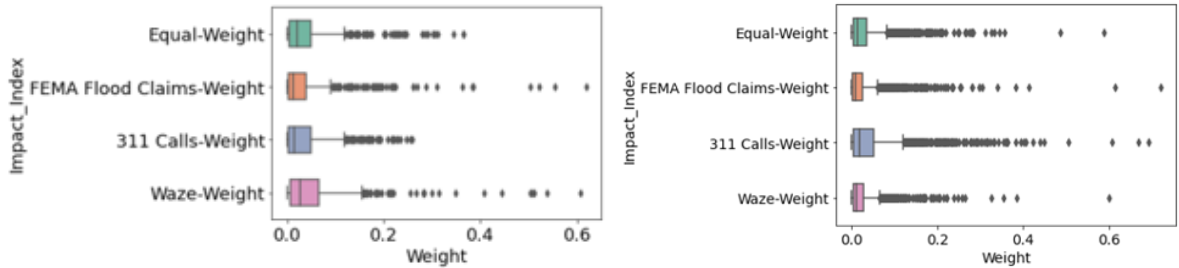

**Fig 2: Box plots for Impact Indexes for Tropical Storm Imelda and Hurricane Ida.**

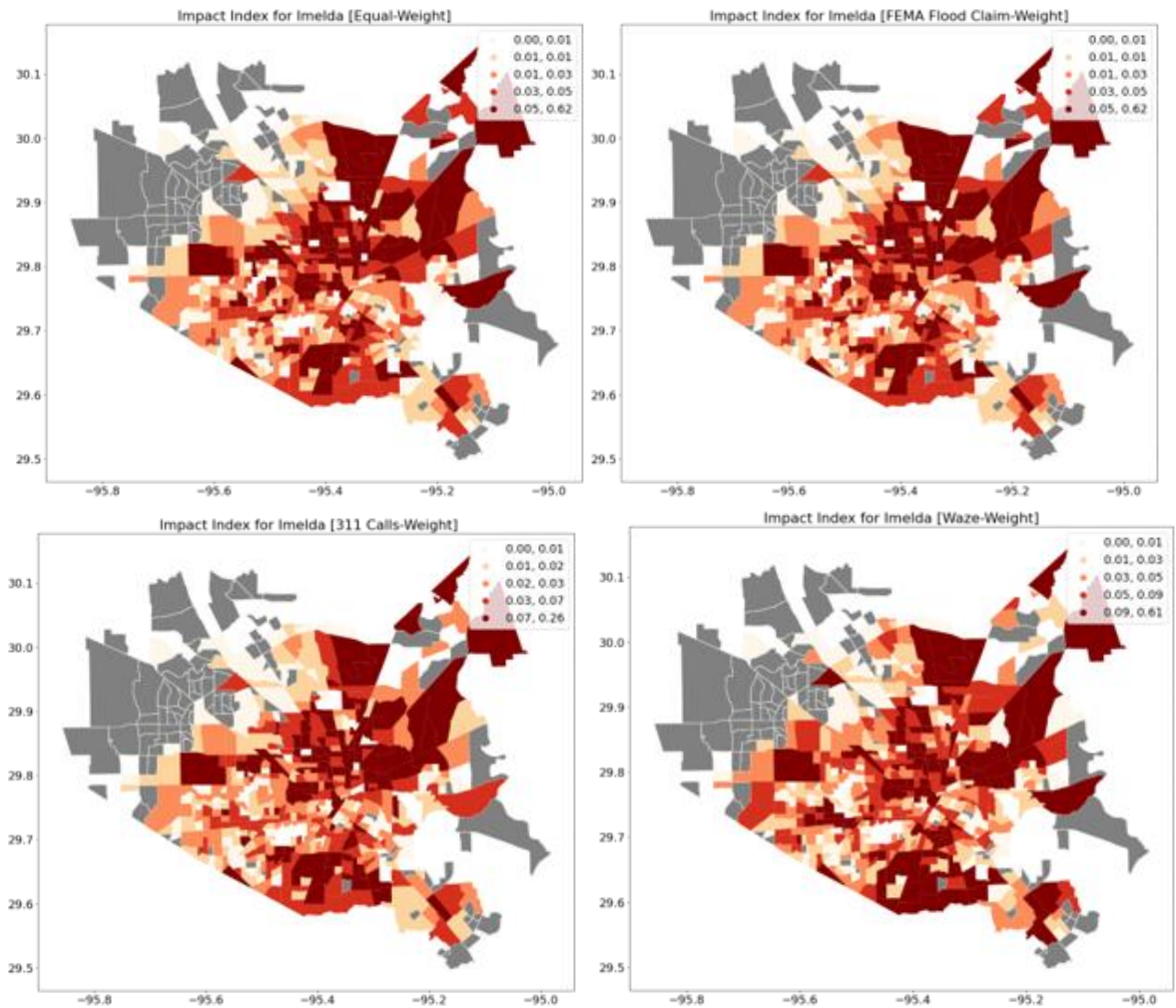

**Fig 3: Spatial Representation of the Quantile Values of Impact Indexes [Equal-Weight and Feature-Weight] For Tropical Storm Imelda in Houston Metropolitan.** The figure shows areas of low values (light orange) and high values (dark red) of the impact indexes. The cutoffs are based on quartile thresholds.

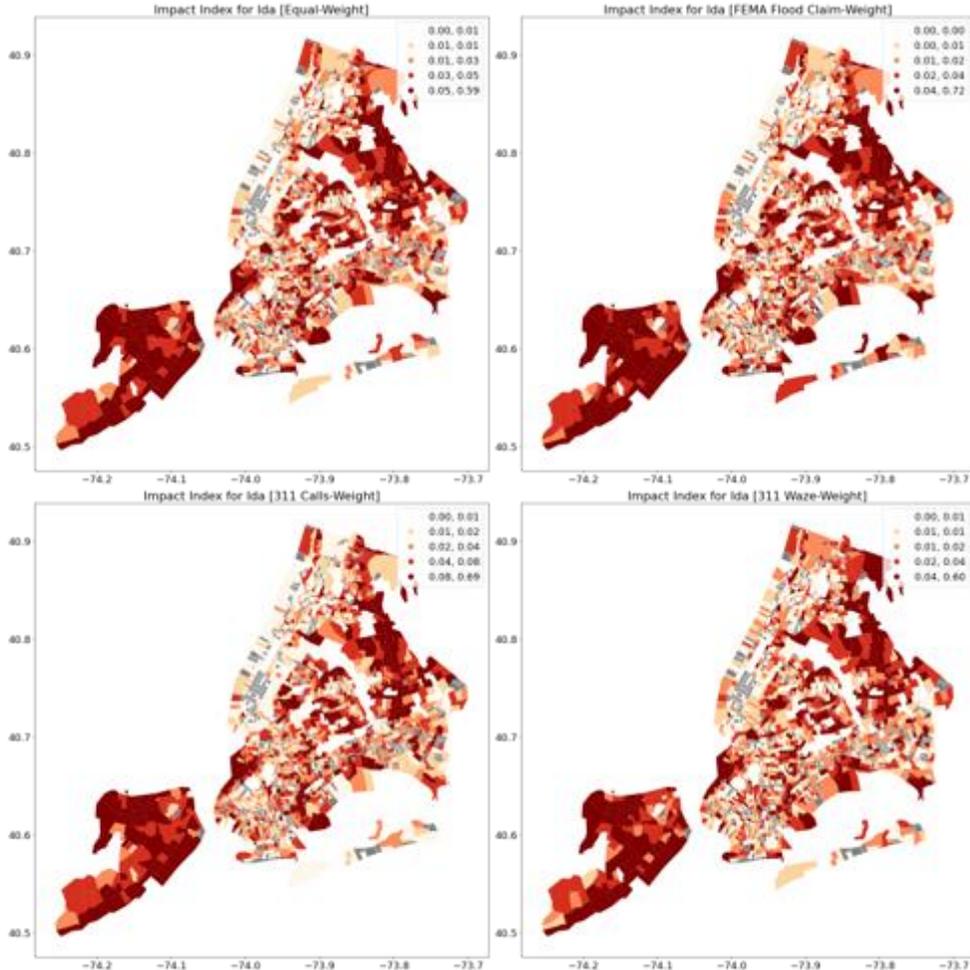

**Fig 4: Spatial Representation of the Quantile Values of Impact Indexes [Equal-Weight and Feature-Weight] For Hurricane Ida in New York City.** The figure shows areas of low values (light orange) and high values (dark red) of the impact indexes. The cutoffs are based on quartile thresholds.

**Vulnerability Clusters:** Using the Elbow Method, we determined the optimal number of clusters for clustering based on sociodemographic features, social capital, physical features, and all variables. K-means algorithm was successfully executed to develop four sets of clusters for each metropolitan area. After clustering each study region based on sociodemographic features, social capital, physical features, and combined variables. Based on the ANOVA results, all variables are shown to be statistically significant ($p < 0.05$) for clustering based on sociodemographic features, social capital, and combined variables. When clustering based on physical features only (POI, road density, and impervious), the POI variable for the Houston Metro Area has a p-value of 0.058. A possible explanation for this is that road density and percentage of impervious cover have a high association, which reduces the effect of POI data on the overall clustering method. Overall, the results establish a significant difference between clusters for the Houston Metropolitan Area and New York City.

As shown in Fig 5 and Fig 6, we created combined vulnerability clusters which accounted for the socio-demographic, social capital, and physical features for Houston and New York City, respectively. In the combined vulnerability cluster of Houston, cluster 0 with the highest impact score was associated with the highest levels of socio-demographic disparities, since cluster 0 had the highest percentage of minority population with 90.55% and lowest per capita income at $19,313. Cluster 3 had the second highest impact score was associated with the highest levels of physical disparities. It had the highest POI count at 0.0871, highest road density at 20.83, and highest percentage of impervious surface at 71.61%. Cluster 3 also had the highest values for bridging and linking social capital, meaning that the affected area could be connected to greater social capital resources from different communities and levels of authority. This cluster was also characterized by the lowest levels of social disparities with the lowest percentage of minority population at 39.71% and the highest per capita income at $73,814. Thus, a combination of disparities must be considered when evaluating for flood risk. Meanwhile, cluster 3 had the highest impact scores. Although it had the second highest physical and socio-demographic disparities, it had the highest social capital values of bonding and linking social capital. Cluster 1 had the second highest impact score, and it was characterized as having the highest percentage of minority population at 87.77%, a lower per capita income at $26,327, and the highest percentage of impervious surface at 81.09%. The k-means clustering analysis allows us to view communities from a high-level perspective as a foundational understanding of the disparities associated with increasing flash-flooding risk.

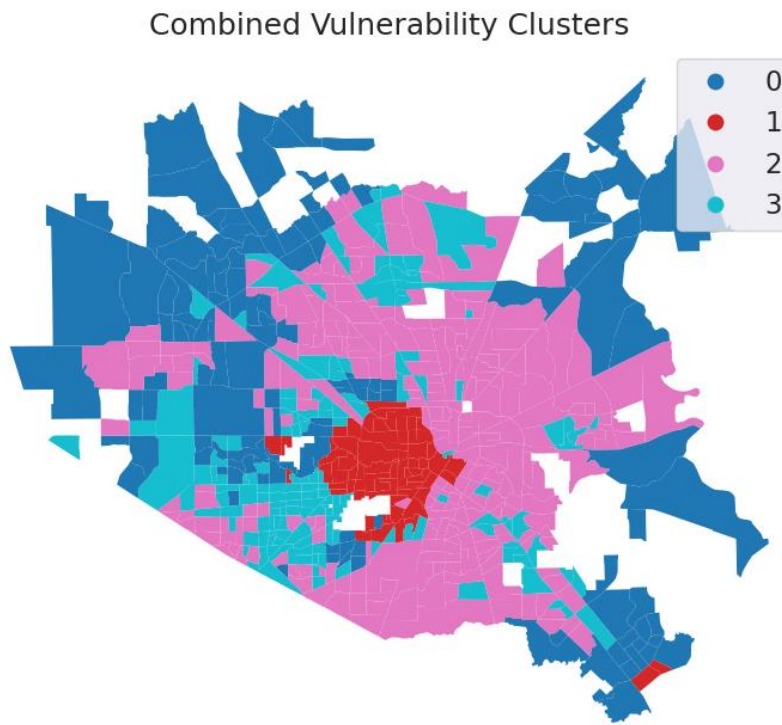

**Fig 5: K-Means Clustering Output of Combined Vulnerability Index for Tropical Imelda in Houston Metropolitan.** The k-means clustering found 4 unique clusters which showed statistically significant ANOVA mean values for impact and the associated features.

**Table 1: Associating features of the combined vulnerability index to the equal-weighted and feature-weighted impact indexes**

| | | 0 | 1 | 2 | 3 |
|---|---|---|---|---|---|
| **Socio-demographic** | % Minority Population | 90.55 | 43.26 | 84.46 | 39.71 |
| | Per Capita Income | 19313.34 | 47720.65 | 21559.59 | 73813.82 |
| | % Elderly | 10.34 | 13.73 | 7.2 | 12.56 |
| | % Person with Disability | 11.28 | 8.85 | 7.92 | 7.51 |
| | % Multi-Unit Housing | 11.10 | 12.81 | 54.14 | 45.41 |
| | Population Density | 4808.15 | 3557.99 | 8824.07 | 7107.65 |
| **Social Capital** | Bonding | 0.4801 | 0.4929 | 0.4259 | 0.4436 |
| | Bridging | 0.4071 | 0.3719 | 0.3918 | 0.6551 |
| | Linking | 0.4745 | 0.5257 | 0.4151 | 0.5602 |
| **Physical** | POI | 0.01236 | 0.03427 | 0.02873 | 0.0871 |
| | Road Density | 16.64 | 13.2 | 14.1 | 20.83 |
| | % Impervious Surface | 54.2 | 46.92 | 66.67 | 71.61 |
| **Impact Indexes** | Equal-Weight | 0.0539 | 0.0242 | 0.0311 | 0.0413 |
| | FEMA Flood Claims-Weight | 0.0497 | 0.0227 | 0.0241 | 0.0289 |
| | 311 Call- Weight | 0.0489 | 0.0235 | 0.0258 | 0.0365 |
| | Waze Report - Weight | 0.0633 | 0.0267 | 0.0434 | 0.0587 |

Combined Vulnerability Cluster

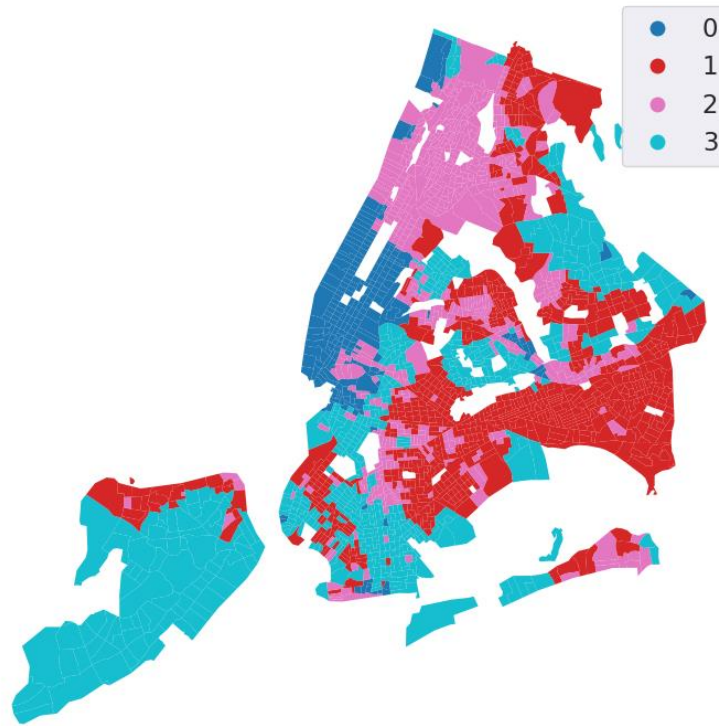

**Fig 6: K-Means Clustering Output for Hurricane Ida in New York City.** The k-means clustering found 4 unique clusters which showed statistically significant ANOVA mean values for impact and the associated features.

**Table 2: Associating features of the combined vulnerability index to the equal-weighted and feature-weighted impact indexes**

| | | 0 | 1 | 2 | 3 |
|---|---|---|---|---|---|
| **Socio-demographic** | % Minority Population | 32.67 | 87.77 | 86.74 | 34.58 |
| | Per Capita Income | 94476.27 | 26327.48 | 23605.04 | 39544 |
| | % Elderly | 18.06 | 13.07 | 12.99 | 15.66 |
| | % Person with Disability | 9.02 | 9.75 | 13.49 | 9.61 |
| | % Multi-Unit Housing | 88.16 | 11.68 | 77.5 | 23.05 |
| | Population Density | 79377.39 | 37404.76 | 76095.54 | 33820.87 |
| **Social Capital** | Bonding | 0.4303 | 0.4591 | 0.4017 | 0.4627 |
| | Bridging | 0.6061 | 0.2242 | 0.24 | 0.2902 |
| | Linking | 0.5016 | 0.601 | 0.5399 | 0.6168 |
| **Physical** | POI | 0.2627 | 0.0169 | 0.029 | 0.0322 |
| | Road Density | 45.38 | 35.00 | 38.76 | 34.08 |
| | % Impervious Surface | 75.91 | 81.09 | 80.22 | 77.16 |
| **Impact Index** | Equal-Weight | 0.0073 | 0.0356 | 0.01439 | 0.0457 |
| | FEMA Flood Claims-Weight | 0.0063 | 0.0261 | 0.0100 | 0.0392 |
| | 311 Call- Weight | 0.009 | 0.0549 | 0.0178 | 0.0649 |
| | Waze Report - Weight | 0.0067 | 0.0259 | 0.0154 | 0.0330 |

**Table 3: ANOVA Testing Results for Combined Vulnerability Cluster**

| | | f-value | p-value | f-value | p-value |
|---|---|---|---|---|---|
| **Socio-demographic** | Percentage of Minority Population | 732.93 | * | 2234.51 | * |
| | Per Capita Income | 373.16 | * | 1050.38 | * |
| | Percentage of Elderly | 51.13 | * | 49.45 | * |
| | Percentage of Person with Disability | 30.44 | * | 96.9 | * |
| | Percentage of Multi-Unit Housing | 343.96 | * | 2808.74 | * |
| | Population Density | 63.744 | * | 326.42 | * |
| **Social Capital** | Bonding | 32.73 | * | 82.44 | * |
| | Bridging | 186.3 | * | 669.9 | * |
| | Linking | 38.98 | * | 78.29 | * |
| **Physical** | POI | 28.49 | * | 117.35 | * |
| | Road Density | 31.84 | * | 93.24 | * |
| | Percentage of Impervious Surface | 83.8 | * | 10.98 | * |
| **Impact Indexes** | Equal-Weight | 14.49 | * | 75.41 | * |
| | FEMA Flood Claim-Weight | 10.08 | * | 62.13 | * |
| | 311 Call- Weight | 16.06 | * | 87.04 | * |
| | Waze Report - Weight | 11.39 | * | 41.05 | * |

*p-value <0.001

## Discussion

Flash flooding causes significant impact to property, critical infrastructure, and human lives. The sudden nature of the flash flooding means that decision-makers must quickly prepare and respond to upcoming threats. Although research has advanced to adequately predict the likelihood and severity of flash flooding, there remain several knowledge gaps to ensure an equitable system for all affected individuals and communities. The application of equitable practices is critical for improving urban community resilience. Thus, the research study proposed a methodology to leverage crowdsourced data and publicly- available datasets to understand the societal impacts of flash flooding events. It uses k-means clustering to associate the socio-demographic, social capital, and physical characteristics of the community.

The research study found several significant findings in the impact index and vulnerability clusters. First, it demonstrated the feasibility of using crowdsourced data to understand the societal impacts of flash flooding. Although each data source has its own limitations, the combination of three crowdsourcing data can show where areas of high impact overlap and identify missed spots of impact. A FEMA flood-claim impact index had the lowest median values of impact score for Houston and New York, which could imply that certain impacts are not fully captured through only FEMA flood claims.

Second, when reviewing the weights of the FEMA Flood Claims, Waze Reports, and 311 calls, it became apparent that the reliability of the crowdsourced data may be location or event specific. The median values and distribution of impact scores were different for each of the impact indexes. Houston had the greatest values from the Waze-weighted impact index while New York City had the greatest values from the 311 calls. The authors speculate that the familiarity and comfortability of the residents may contribute to these differing scores. Houstonians may be more inclined to drive their own vehicles and use transportation-related apps while New Yorkers may use public transportation or other pedestrian methods of traveling.

Third, the inconsistency in reporting from crowdsourced data can influence the reliability of reporting. As mentioned in the methodology, the 311 call system from Houston had a "Flood" report available. However, the 'Storm' and 'Sewer' reports needed to be used for New York. In order for crowdsourcing data to be successful, both individuals and the community leaders must be fully aware of its potential.

Fourth, the clusters show that a combination of socio-demographic, social capital, and physical disparities may be associated with higher impact index. In Houston, the cluster with the highest impact index score also had the highest percentage of minority and the lowest per capita income while the cluster with the second highest impact index score had the highest percentage of impervious surface, highest road density, and highest POI count. In New York, the cluster with the second highest impact index score had both the highest percentage of minority and highest percentage of impervious surfaces. Thus, the findings support the need to further explore the interconnections between physical and social vulnerabilities to flash flooding.

**Concluding Remarks**

The results suggest that a combination of social and physical features may be related to a community's level of flash flooding vulnerability. This is revealed through a difference in flash flood impact amongst k-means clusters. The findings not only indicate a significant difference between clusters defined on the basis of socio-demographic, social capital, and physical features, but also show a statistically significant difference in the flash flooding impact experienced across each cluster. This supports the proposition that publicly available data quantifying the socio-demographic, social capital, and physical characteristics of a region can be used as a resource to better understand flash flood vulnerability. Thus, the extent to which hotspots for flash floods share similar social and physical features is significant. However, further research is needed to understand how each measured variable specifically contributes to flash flood vulnerability and impact.

The research focus was on ensuring that municipalities and agencies could all understand and operate the flash flooding impact index. A comprehensive understanding of the multiple variables which form the impact index and how they are used to represent flash flood impact is essential for all involved agencies. Furthermore, it is important for local authorities and decision-makers to know how to interpret the impact indices and how they relate to the area's social and physical features. The k-means unsupervised learning algorithm served as a foundational start to understand associations between flash flooding impact and socio-demographic, social capital, and physical features. For instance, in Houston, factors like percentage of minority and income were most significant, while in New York City, percentage of minority and impervious surface were more prominent. Further incorporation of machine learning techniques and other relevant features could enhance our understanding of the factors influencing flash flooding impact.

Due to the potential of crowdsourcing data, the authors recommend further investment in integrating crowdsourcing data for quick-time emergency decisions. This can be pursued through establishing and expanding partnerships with the Data for Good Initiative or by offering enhanced data analytics training within municipalities to foster meaningful and data-driven decision-making. Decision makers must also understand the appropriateness of using certain data in their community. This could be done by selecting the right crowdsourcing and adjusting the weights according to the community preferences. A community that is not familiar with the Waze app would be misrepresented in a Waze-weighted impact index. For example, we assume Houstonians are more familiar with using Waze than New Yorkers, given Houston's higher car usage compared to New York City's robust public transit and dense urban setting. Residents may also feel more comfortable and more aware of certain crowdsourcing reporting. This was evident in the differences in 311 reporting between the two cities analyzed: Houston residents reported

'Flooding' and 'Drainage' issues during Tropical Storm Imelda, whereas New York City residents reported 'Storm' and 'Sewer' complaints during Hurricane Ida. Building community trust may also encourage residents to use crowdsourced data that are perceived to relate to authoritative and government entities. The impact-index is a foundational step in incorporating societal impacts in flash flood models and assisting in the equitable distribution of resources.

Future research could further explore the inherent biases of using crowdsourced data and how to mitigate such biases in theory and in practice. Factors such as socioeconomic status, race, and age can significantly impact who is represented in digital datasets, leading to potential biases. Therefore, it's crucial to consider and critically assess the comprehensiveness of data sources, recognizing that many groups, such as digitally invisible populations, are likely to be underrepresented. Additionally, biases in population representation could stem from individuals' comfort level with sharing information with government sources, the accessibility and affordability of phones, and the expectation of reporting on individuals. Advances in data, such as location intelligence and human mobility data, can be used to supplement crowdsourced data sources. Location intelligence, including behavioral, economic, and demographic data, can be overlaid on crowdsourced data to provide spatial context and better understand how different populations are affected by natural hazards. Human mobility data can be used to validate crowdsourced data and improve the accuracy of predictive models. For example, crowdsourced data on damage can be augmented with location and mobility data to better predict high-impact areas and allocate resources more effectively.

Through a deeper understanding of how similar social and physical features contribute to the development of flash flood hotspots, urban areas can implement more effective land-use strategies and infrastructural improvements, direct flood mitigation resources and efforts towards socio-demographic groups which are disproportionately affected by flash floods, and develop more effective emergency response plans. This understanding guides city planners in making informed decisions about where to allow development and how to design flood-resistant infrastructure. Insights into the physical features contributing to flash flood hotspots can inform infrastructure improvements, such as upgraded drainage systems or flood barrier construction. Furthermore, more effective flash flood mitigation efforts, including community engagement and education, can be developed by understanding the socio-demographic characteristics that define flood risk. Emergency response plans can be targeted to meet the specific needs of affected communities. Moreover, policymakers can leverage this data-driven understanding of flash flood risk to allocate funding and resources more strategically.

- The research leveraged three sources of crowdsourced data to create a flash flood impact index. The results support the use of publicly available data to understand flood vulnerability while also emphasizing the need for further research into the specific contributions of each variable.
- The study demonstrates the association between flash flood vulnerability and social and physical features, as indicated by significant differences in flash flood impacts across k-means clusters based on socio-demographic, social capital, and physical characteristics. It is essential for all involved agencies to understand regional flood characteristics and how they relate to social and physical features. Additionally, the results underscore the varying significance of socio-demographic factors like percentage of minority and income in different urban contexts, such as Houston and New York City, indicating the necessity for region-specific flood risk management strategies.

- Decision makers should explore the appropriateness of different crowdsourced data sources, such as Waze reports and 311 calls, and adjust the impact index weights according to community preferences.
- Future research is encouraged to explore and address the biases in crowdsourced data, such as underrepresentation of digitally invisible populations. Augmentation of crowdsourced data with location intelligence and human mobility data may be used for more comprehensive and accurate flash flood impact assessments.
- An understanding of how social and physical features interact to form flash flood hotspots can be used to guide policymakers and urban planners in developing targeted land-use strategies, infrastructural improvements, and emergency response plans, thereby ensuring equitable resource distribution and enhancing community resilience against flash floods.

**Data Availability Statement:** Some or all data, models, or code generated or used during the study are proprietary or confidential in nature and may only be provided with restrictions. Open-source data which are publicly available have been cited in the article. Please contact the authors if you are interested in accessing the data.

**Acknowledgments:** This material is based in part upon work supported by the National Science Foundation, USA under the CRISP 2.0 Type 2 No. 1832662 grant and National Science Foundation Graduate Research Fellowship. The authors also would like to acknowledge the data support from Waze. Any opinions, findings, conclusions, or recommendations expressed in this material are those of the authors and do not necessarily reflect the views of the National Science Foundation and Waze.

**Author Contributions:** All authors critically revised the manuscript, gave final approval for publication, and agree to be held accountable for the work performed therein. Allison Clarke and Natalie Coleman are co-first authors, who were responsible for guiding data collection, performing analysis, interpreting the significant results, and writing the manuscript. Dr. Ali Mostafavi was a faculty advisor for the project and provided critical feedback on the literature review development, analysis and manuscript.

**References:**

Abdel-Mooty MN, Yosri A, El-Dakhakhni W, et al. (2021) Community flood resilience categorization framework. *International Journal of Disaster Risk Reduction* 61: 102349.

Aldrich DP (2010) Fixing recovery: Social capital in post-crisis resilience. *Journal of Homeland Security, Forthcoming*.

Aldrich DP and Meyer MA (2015) Social capital and community resilience. *American behavioral scientist* 59(2): 254-269.

Alipour A, Ahmadalipour A and Moradkhani H (2020) Assessing flash flood hazard and damages in the southeast United States. *Journal of Flood Risk Management* 13(2): e12605.

Azad MJ and Pritchard B (2023) Bonding, bridging, linking social capital as mutually reinforcing elements in adaptive capacity development to flood hazard: Insights from rural Bangladesh. *Climate Risk Management* 40: 100498.

Blum AG, Ferraro PJ, Archfield SA, et al. (2020) Causal effect of impervious cover on annual flood magnitude for the United States. *Geophysical Research Letters* 47(5): no-no.

Centers for Disease Control and Prevention (2018) Social Vulnerability Index.

Chacowry A, McEwen LJ and Lynch K (2018) Recovery and resilience of communities in flood risk zones in a small island developing state: A case study from a suburban settlement of Port Louis, Mauritius. *International Journal of Disaster Risk Reduction* 28: 826-838.

Chakraborty J, Collins Timothy W, Montgomery Marilyn C, et al. (2014) Social and Spatial Inequities in Exposure to Flood Risk in Miami, Florida. *Natural Hazards Review* 15(3): 04014006.

Chakraborty L, Rus H, Henstra D, et al. (2020) A place-based socioeconomic status index: Measuring social vulnerability to flood hazards in the context of environmental justice. *International Journal of Disaster Risk Reduction* 43: 101394.

Coleman N, Esmalian A and Mostafavi A (2020) Anatomy of susceptibility for shelter-in-place households facing infrastructure service disruptions caused by natural hazards. *International Journal of Disaster Risk Reduction* 50: 101875.

Cutter Susan L, Emrich Christopher T, Gall M, et al. (2018) Flash Flood Risk and the Paradox of Urban Development. *Natural Hazards Review* 19(1): 05017005.

Davis M (2022) *Estimated $29.4 Billion in Property Damage from Severe Weather Not Covered by Insurance in Past 5 Years.* . Available at: https://www.valuepenguin.com/severe-weather-property-damages-study.

Działek J, Biernacki W and Bokwa A (2014) Impact of social capital on local communities' response to floods in southern Poland. *Risks and conflicts: Local responses to natural disasters*. Emerald Group Publishing Limited, pp.185-205.

Emrich CT, Aksha SK and Zhou Y (2022) Assessing distributive inequities in FEMA's Disaster recovery assistance fund allocation. *International Journal of Disaster Risk Reduction* 74: 102855.

Emrich CT, Tate E, Larson SE, et al. (2020) Measuring social equity in flood recovery funding. *Environmental Hazards* 19(3): 228-250.

Esparza M, Farahmand H, Brody S, et al. (2023) Examining data imbalance in crowdsourced reports for improving flash flood situational awareness. *International Journal of Disaster Risk Reduction* 95: 103825.

Faber JW (2015) Superstorm Sandy and the demographics of flood risk in New York City. *Human Ecology* 43: 363-378.

Fan C, Esparza M, Dargin J, et al. (2020) Spatial biases in crowdsourced data: Social media content attention concentrates on populous areas in disasters. *Computers, Environment and Urban Systems* 83: 101514.

Farahmand H, Liu X, Dong S, et al. (2022) A network observability framework for sensor placement in flood control networks to improve flood situational awareness and risk management. *Reliability Engineering & System Safety* 221: 108366.

Feng B, Zhang Y and Bourke R (2021) Urbanization impacts on flood risks based on urban growth data and coupled flood models. *Natural Hazards* 106(1): 613-627.

Fernandez P, Mourato S, Moreira M, et al. (2016) A new approach for computing a flood vulnerability index using cluster analysis. *Physics and Chemistry of the Earth, Parts a/b/c* 94: 47-55.

Fernandez S (2019) Tropical Storm Imelda's flooding turns deadly as southeast Texas is swamped by rain. *The Texas Tribune*.

Fussell E, Sastry N and VanLandingham M (2010) Race, socioeconomic status, and return migration to New Orleans after Hurricane Katrina. *Population and Environment* 31(1): 20-42.

Henao Salgado MJ and Zambrano Nájera J (2022) Assessing flood early warning systems for flash floods. *Frontiers in Climate* 4: 787042.

Hudson P, Hagedoorn L and Bubeck P (2020) Potential Linkages Between Social Capital, Flood Risk Perceptions, and Self-Efficacy. *International Journal of Disaster Risk Science* 11(3): 251-262.

Jones JA, Swanson FJ, Wemple BC, et al. (2000) Effects of Roads on Hydrology, Geomorphology, and Disturbance Patches in Stream Networks. *Conservation Biology* 14(1): 76-85.

Karagiorgos K, Thaler T, Heiser M, et al. (2016) Integrated flash flood vulnerability assessment: Insights from East Attica, Greece. *Journal of Hydrology* 541: 553-562.

Kontokosta CE and Hong B (2021) Bias in smart city governance: How socio-spatial disparities in 311 complaint behavior impact the fairness of data-driven decisions. *Sustainable Cities and Society* 64: 102503.

Kontokosta CE and Malik A (2018) The Resilience to Emergencies and Disasters Index: Applying big data to benchmark and validate neighborhood resilience capacity. *Sustainable Cities and Society* 36: 272-285.

Kraft A and Usbeck R (2022) The Ethical Risks of Analyzing Crisis Events on Social Media with Machine Learning. *arXiv preprint arXiv:2210.03352*.

Latto A and Berg R (2020) Tropical Storm Imelda. Reportno. Report Number|, Date. Place Published|: Institution|.

Lee C-C, Maron M and Mostafavi A (2022) Community-scale big data reveals disparate impacts of the Texas winter storm of 2021 and its managed power outage. *Humanities and Social Sciences Communications* 9(1): 1-12.

Lee NT (2024) *Digitally Invisible: How the Internet Is Creating the New Underclass.* Brookings Institution Press.

Li B, Fan C, Chien Y-H, et al. (2023) Mobility behaviors shift disparity in flood exposure in us population groups. *arXiv preprint arXiv:2307.01080*.

Lin C-Y, Chen W-C, Chang P-L, et al. (2011) Impact of the Urban Heat Island Effect on Precipitation over a Complex Geographic Environment in Northern Taiwan. *Journal of Applied Meteorology and Climatology* 50(2): 339-353.

Lin K, Chen H, Xu C-Y, et al. (2020) Assessment of flash flood risk based on improved analytic hierarchy process method and integrated maximum likelihood clustering algorithm. *Journal of Hydrology* 584: 124696.

Liu J, Xiong J, Chen Y, et al. (2023) A new avenue to improve the performance of integrated modeling for flash flood susceptibility assessment: applying cluster algorithms. *Ecological Indicators* 146: 109785.

Liu Z, Felton T and Mostafavi A (2024) Interpretable machine learning for predicting urban flash flood hotspots using intertwined land and built-environment features. *Computers, Environment and Urban Systems* 110: 102096.

Lo AY and Chan F (2017) Preparing for flooding in England and Wales: the role of risk perception and the social context in driving individual action. *Natural Hazards* 88(1): 367-387.

Longo J, Kuras E, Smith H, et al. (2017) Technology Use, Exposure to Natural Hazards, and Being Digitally Invisible: Implications for Policy Analytics. *Policy & Internet* 9(1): 76-108.

Lowrie C, Kruczkiewicz A, McClain SN, et al. (2022) Evaluating the usefulness of VGI from Waze for the reporting of flash floods. *Scientific reports* 12(1): 5268.

Lu XS and Zhou M Analyzing the evolution of rare events via social media data and k-means clustering algorithm. IEEE, 1-6.

Marcius CR and Norman B (2021) They Put Everything Into Their Homes. Not One Was Spared in the Flood. *The New York Times*.

Masozera M, Bailey M and Kerchner C (2007) Distribution of impacts of natural disasters across income groups: A case study of New Orleans. *Ecological Economics* 63(2): 299-306.

McAllister T and McAllister T (2013) *Developing guidelines and standards for disaster resilience of the built environment: A research needs assessment.* US Department of Commerce, National Institute of Standards and Technology ….

McKinley J, Rubinstein D and Mays J (2021a) The Storm Warnings Were Dire. Why Couldn't New York Be Protected? *The New York Times*.

McKinley J, Schweber N, Rosa A, et al. (2021b) New York Flooding: Flooding From Ida Kills Dozens of People in Four States, in New York Times. . *The New York Times*.

Mergel I, Rethemeyer RK and Isett K (2016) Big Data in Public Affairs. *Public Administration Review* 76(6): 928-937.

Mobley W and Blessing R (2022) Using machine learning to predict flood hazards based on historic damage. *Coastal Flood Risk Reduction*. Elsevier, pp.61-75.

Mobley W, Sebastian A, Blessing R, et al. (2021) Quantification of continuous flood hazard using random forest classification and flood insurance claims at large spatial scales: a pilot study in southeast Texas. *Natural hazards and earth system sciences* 21(2): 807-822.

Mobley W, Sebastian A, Highfield W, et al. (2019) Estimating flood extent during Hurricane Harvey using maximum entropy to build a hazard distribution model. *Journal of Flood Risk Management* 12: e12549.

Mohtar WHMW, Abdullah J, Maulud KNA, et al. (2020) Urban flash flood index based on historical rainfall events. *Sustainable Cities and Society* 56: 102088.

Multi-Resolution Land Characteristics and United States Geological Survey (USGS) (2021) Land Cover and Land Use.

Nainggolan R, Perangin-angin R, Simarmata E, et al. (2019) Improved the Performance of the K-Means Cluster Using the Sum of Squared Error (SSE) optimized by using the Elbow Method. *Journal of Physics: Conference Series* 1361(1): 012015.

Napieralski JA and Giroux B (2019) Quantifying Proximity and Conformity between Road Networks, Urban Streams, and Watershed Boundaries. *Annals of the American Association of Geographers* 109(1): 35-49.

National Academies of Sciences E and Medicine (2019) *Framing the Challenge of Urban Flooding in the United States.* Washington, DC: The National Academies Press.

National Oceanic and Atmospheric Administration (NOAA) *What is Flash Flooding*. Available at: https://www.weather.gov/phi/FlashFloodingDefinition.

Negri R, Tsai Y-JJ, Tan BY, et al. Investigating the Use of Citizen-Science Data as a Proxy for Flood Risk Assessment in New York City.

New York City (2021) Open Data in New York City,.

Open Data Portal in Houston (2019) Open Data Portal. In: Houston Co (ed).

Overton M, Larson S, Carlson L, et al. (2022) Public data primacy: the changing landscape of public service delivery as big data gets bigger. *Innovations Technology Governance Globalization* 2.

Patrascu FI, Mostafavi A and Vedlitz A (2023) Disparities in access and association between access to critical facilities during day-to-day and disrupted access as a result of storm extreme weather events. *Heliyon* 9(8).

Praharaj S, Zahura FT, Chen TD, et al. (2021) Assessing trustworthiness of crowdsourced flood incident reports using Waze data: A Norfolk, Virginia case study. *Transportation research record* 2675(12): 650-662.

Rachma ST and Lin Y-C (2023) Investigating the changing spatiotemporal urban heat island and its impact on thunderstorm patterns by Hilbert Huang transform. *Stochastic Environmental Research and Risk Assessment*. DOI: 10.1007/s00477-023-02571-5.

Rahman M, Ningsheng C, Mahmud GI, et al. (2021) Flooding and its relationship with land cover change, population growth, and road density. *Geoscience Frontiers* 12(6): 101224.

Rainey JL, Pandian K, Sterns L, et al. (2021) Using 311-Call data to Measure Flood Risk and Impacts: The Case of Harris Country TX. *Institute for a Disaster Resilient Texas: Galveston, TX, USA* 22.

Ramadhan MI An analysis of natural disaster data by using K-means and K-medoids algorithm of data mining techniques. IEEE, 221-225.

Robertson I, A. Bobet, B. Edge, W. Holmes, D. Johnson, M. LaChance, J. Ramirez. (2023) *Natural Hazards Engineering Research Infrastructure, Science Plan, Multi-Hazard Research to Make a More Resilient World, Third Edition*. Available at: https://doi.org/10.17603/ds2-abbs-0966.

Rufat S, Tate E, Burton CG, et al. (2015) Social vulnerability to floods: Review of case studies and implications for measurement. *International Journal of Disaster Risk Reduction* 14: 470-486.

Rustinsyah R, Prasetyo RA and Adib M (2021) Social capital for flood disaster management: Case study of flooding in a village of Bengawan Solo Riverbank, Tuban, East Java Province. *International Journal of Disaster Risk Reduction* 52: 101963.

Safaei-Moghadam A, Tarboton D and Minsker B (2023) Estimating the likelihood of roadway pluvial flood based on crowdsourced traffic data and depression-based DEM analysis. *Natural hazards and earth system sciences* 23(1): 1-19.

SafeGraph (2020) Points-of-Interest Data.

Sanders BF, Schubert JE, Kahl DT, et al. (2023) Large and inequitable flood risks in Los Angeles, California. *Nature Sustainability* 6(1): 47-57.

Sjöstrand K (2022) Urbanization impacts on floods. *Nature Reviews Earth & Environment* 3(11): 738-738.

Sohn W, Kim J-H, Li M-H, et al. (2020) How does increasing impervious surfaces affect urban flooding in response to climate variability? *Ecological Indicators* 118: 106774.

Špitalar M, Gourley JJ, Lutoff C, et al. (2014) Analysis of flash flood parameters and human impacts in the US from 2006 to 2012. *Journal of Hydrology* 519: 863-870.

Supriyadi B, Windarto AP and Soemartono T (2018) Classification of natural disaster prone areas in Indonesia using K-means. *International Journal of Grid and Distributed Computing* 11(8): 87-98.

Tammar A, Abosuliman SS and Rahaman KR (2020) Social capital and disaster resilience nexus: a study of flash flood recovery in Jeddah City. *Sustainability* 12(11): 4668.

Tang S, Shu X, Shen S, et al. (2017) Study of portable infrastructure-free cell phone detector for disaster relief. *Natural Hazards* 86(1): 453-464.

Tate E, Rahman MA, Emrich CT, et al. (2021) Flood exposure and social vulnerability in the United States. *Natural Hazards* 106(1): 435-457.

Taylor DB and Mervosh S (2019) Imelda Hits Houston Area With Rain and Threatens to Bring More. *The New York Times*.

Tellman B, Schank C, Schwarz B, et al. (2020) Using disaster outcomes to validate components of social vulnerability to floods: Flood deaths and property damage across the USA. *Sustainability* 12(15): 6006.

Texas Department of Transportation Open Data Portal.

Waze (2019, 2021) Waze Traffic Data.

Xu H, Ma C, Lian J, et al. (2018) Urban flooding risk assessment based on an integrated k-means cluster algorithm and improved entropy weight method in the region of Haikou, China. *Journal of Hydrology* 563: 975-986.

Yuan F, Farahmand H, Blessing R, et al. (2023) Unveiling dialysis centers' vulnerability and access inequality during urban flooding. *Transportation Research Part D: Transport and Environment* 125: 103920.

Yuan F and Liu R (2018) Feasibility study of using crowdsourcing to identify critical affected areas for rapid damage assessment: Hurricane Matthew case study. *International Journal of Disaster Risk Reduction* 28: 758-767.

Zhang R, Chen Y, Zhang X, et al. (2022) Mapping homogeneous regions for flash floods using machine learning: A case study in Jiangxi province, China. *International Journal of Applied Earth Observation and Geoinformation* 108: 102717.